# Quantization as a Consequence of Symmetry

J. Towe

Department of Physics, The Antelope Valley College
Lancaster, CA, USA
jtowe@avc.edu

It is argued that the Heisenberg relation on 4-spacetime is a necessary condition for the local gauge invariance of a classical wave in the 5-dimensional Kaluza-Klein theory.

# The Kaluza-Klein Theory, Gauge Invariance and Quantization

Let us recall the Christoffel connections (of the second kind) that underlie general relativity mechanics on 4-spacetiime:

$$\left\{ \begin{matrix} \mu \\ \nu \quad \rho \end{matrix} \right\} = \frac{g^{\mu\alpha}}{2} \left( \frac{\partial g_{\alpha\nu}}{\partial x^{\rho}} + \frac{\partial g_{\rho\alpha}}{\partial x^{\nu}} - \frac{\partial g_{\nu\rho}}{\partial x^{\alpha}} \right): \qquad (1)$$

μ, ν, ρ = 1, 2, 3, 4 [A. Einstein, 1915]. An analogous set of connection coefficients characterize the general theory of relativity on 5-spacetime that was considered by Theodor Kaluza and Oscar Klein [T. Kaluza, 1921; O. Klein, 1926; O. Klein, 1938]:

$$\left\{ \begin{matrix} M \\ N \quad R \end{matrix} \right\} = \frac{g^{MA}}{2} \left( \frac{\partial g_{AN}}{\partial x^{R}} + \frac{\partial g_{RA}}{\partial x^{N}} - \frac{\partial g_{NR}}{\partial x^{A}} \right): \qquad (2)$$

M, N, R = 1, 2, 3, 4, 5. The coefficients

$$\left\{ \begin{matrix} \mu \\ \nu \quad 5 \end{matrix} \right\} = \frac{g^{\mu\alpha}}{2} \left( \frac{\partial g_{\alpha\nu}}{\partial x^{5}} + \frac{\partial g_{5\alpha}}{\partial x^{\nu}} - \frac{\partial g_{\nu 5}}{\partial x^{\alpha}} \right): \qquad (3)$$

μ,ν = 1, 2, 3, 4 clearly describe the connection coefficients that are, in addition to those described by (1), perceived on 4-spacetime in the context of the Kaluza-Klein theory. The 'Kaluza-Klein' condition

$$\frac{\partial g_{\alpha\nu}}{\partial x^{5}} = 0 \qquad (4)$$

reduces the coefficients (3) to

$$F^{\mu}{}_{\nu} = \frac{g^{\mu\alpha}}{2} \left( \frac{\partial g_{5\alpha}}{\partial x^{\nu}} - \frac{\partial g_{\nu 5}}{\partial x^{\alpha}} \right), \qquad (5)$$

which Kaluza and Klein interpreted as the electromagnetic field tensor on 4-spacetime. Let us now consider the energy, $\mathcal{L}$, of this electromagnetic field in the presence of 4-current. If the energy, $\mathcal{L}$, is described in terms of $F^{\mu}{}_{\nu}$, if $\mathcal{L}$ is invariant under the group U(1) of 1-parameter phase transformations, and if the variational principle

$$\delta \int \mathcal{L} \, d^{4}x = 0,$$

is restricted to a variation of action in terms of the metrical coefficients $g_{\mu 5}$, then the Euler-Lagrange equations are, in the Kaluza-Klein model, the Maxwell equations (that describe electromagnetic interactions in 4-spacetime). If these are integrated, the result is a wave equation describing the propagation of a classical electromagnetic wave in 4-spacetime.

A second result of the projection that is required by (4) is the reduction

$$GL(5) \rightarrow GL(4) \times U(1),$$

where GL(5) and GL(4) respectively represent the groups of general rotations on 5-spacetime and 4-spacetime, under which the equations of the 5-dimensioinal theory of general relativity and the 4-dimensional theory of general relativity are invariant; and where U(1) represents the above described group of 1-parameter phase transformations.

The sets GL(N) and U(1) of transformations are called 'groups' because the rules that are satisfied by the multiplicative combinations of the exponential transformations include

1. associativity
and
2. commutativity (if abelian group),

and the elements themselves include

3. a multiplicative identity
and
4. a multiplicative inverse.

The elements $\exp(\sum_{\alpha=1}^{N(N-1)/2} G_\alpha u^\alpha)$ that constitute the general linear group GL(N) can be expanded about the identity to yield

$$I + \sum_{\alpha=1}^{N(N-1)/2} G_\alpha u^\alpha + \ldots,$$

where $u^\alpha$ and $G_\alpha$ respectively represent the parameters and generators of the GL(N) group. To first order in the parameters $u^\alpha$, the above expansion reduces to

$$I + \sum_{\alpha=1}^{N(N-1)/2} G_\alpha u^\alpha.$$

GL(N) is said to be a continuous group because the terms $\sum_{\alpha=1}^{N(N-1)/2} G_\alpha u^\alpha$ can be infinitesimal (so that expansion about the identity is continuous).

The number of parameters N(N-1)/2 is obtained by taking the difference between the number of variables to be determined and the number of degrees of freedom that are involved in the transformation equations (for flat; i.e. Minkowski spacetime, these transformation equations are the Lorentz transformations). Thus, N(N-1)/2 is the number of parameters that is required to uniquely determine the group GL(N). The action of

GL(4) is on 4-momentum, so that, by Noether's theorem, the invariance of an interaction under GL(4) is equivalent to the conservation, by that interaction, of 4-momentum.

The group U(1) consists (to first order in the parameter $u = u_\mu dx^\mu$) of the infinitesimal phase transformations

$$I + \sum_{\mu=1}^{4} u_\mu dx^\mu,$$

U(1) is usually parameterized in terms of electrical charge (so that the invariance of an interaction under U(1) is, by Noether's theorem, equivalent to the conservation by the interaction of electrical charge), but this is not required.

Now, adopting the Einstein summation convention; i.e. summing over repeated Greek indices from 1 through 4, let us consider a classical electromagnetic wave in the Kaluza-Klein theory; and the transformation of this wave under U(1) (generic parameter):

$$\hat{\Psi}^\alpha = \Psi^\alpha \expi \frac{u_\mu dx^\mu}{k}, \tag{6}$$

where k is the smallest value with the dimensions of $u = \sum_{\mu=1}^{4} u_\mu dx^\mu$.

The invariance of this classical wave state under the group U(1) clearly implies that

$$\hat{\Psi}^\alpha = \Psi^\alpha \expi \frac{u_\mu dx^\mu}{k} = \Psi^\alpha \expi(2\pi n) = \Psi^\alpha \cos(2\pi n) = \Psi^\alpha, \tag{7}$$

where n is an integer, or that

$$\frac{u_\mu dx^\mu}{k} = 2\pi n; \tag{8}$$

which, due to arithmetic considerations, restricts k to a non-zero value. If U(1) is parameterized in terms of the '4-momentum 1-form' then, (8) implies that

$$\frac{P_\mu dx^\mu}{k} = 2\pi n: \tag{9}$$

$P_4 = E/ic$, $dx^4 = icdt$, where k assumes the dimensions of action. Because negative values of action are not considered, expression 9 restricts k to a positive value: n = 1,2,3,… Thus

$$P_\mu dx^\mu \geq 2\pi k. \tag{10}$$

If axes are aligned so that motion is parallel to the μ-axis then, one can write

$$P_\mu dx^\mu \geq 2\pi k, \tag{11}$$

where summation is not implied. Thus, since every axis can be aligned with motion, one can write

$$P_\mu dx^\mu \geq 2\pi k: \mu = 1, 2, 3, 4. \tag{12}$$

Finally, since k has the dimensions of action, one can assign to k a constant value:

$$k = \frac{\hbar}{4\pi}, \tag{13}$$

so that (11) reduces to

$$P_\mu dx^\mu \geq \frac{\hbar}{2}: \mu = 1, 2, 3, 4; \tag{14}$$

which are the Heisenberg relations on 4-spacetime [L. Schiff, 1968].

## Conclusion

The notion that quantum mechanics can emerge from classical theory [J. Towe, 1988]; e.g. from general relativity is often described as infeasible due to the counterintuitive nature of quantum theory. Even though quantum physics contains less information than classical physics, it is argued that the many strange and counterintuitive concepts that are involved in the former can hardly be accounted for in terms of classical physics. However, it should be emphasized that one in not obligated to account for the details of quantum theory. All of quantum mechanics (up to relativistic considerations) can be derived from the Heisenberg uncertainty principle, or from the logically equivalent Schrodinger formulation. The essential element is simply the non-zero value of the quantum of action.

.